\documentclass[letter,twocolumn]{jpsj2} 
\title{Softening of Magnetic Excitations Leading to Pressure-Induced Quantum Phase Transition in Gapped Spin System KCuCl$_3$}

\author{Kenji {\sc Goto}\thanks{E-mail: goto.kenji@jaea.go.jp.}$^{1,4}$, Toyotaka \textsc{Osakabe}$^{1}$, Kazuhisa {\sc Kakurai}\thanks{E-mail: kakurai.kazuhisa@jaea.go.jp.}$^{1}$, Yoshiya {\sc Uwatoko}$^{2}$,\\ Akira {\sc Oosawa}$^{3}$,  
Jun {\sc Kawakami}$^{4}$ and Hidekazu {\sc Tanaka}$^{4}$}

\inst{$^{1}$Quantum Beam Science Directorate, Japan Atomic Energy Agency, Tokai, Ibaraki 319-1195\\
$^{2}$Institute for Solid State Physics, The University of Tokyo, 
Kashiwanoha, Kashiwa, Chiba 277-8581 \\
$^{3}$Department of Physics, Sophia University, Kioi-cho, Chiyoda-ku, Tokyo 102-8554 \\
$^{4}$Department of Physics, Tokyo Institute of Technology, Oh-okayama, Meguro-ku, Tokyo 152-8551}

\abst{KCuCl$_3$ is a three dimensionally coupled spin dimer system, which undergoes a pressure-induced quantum phase transition from a gapped ground state to an antiferromagnetic state at a critical pressure of $P_{\rm c} \simeq 8.2$ kbar. Magnetic excitations in KCuCl$_3$ at a hydrostatic pressure of 4.7 kbar have been investigated by conducting neutron inelastic scattering experiments using a newly designed cylindrical high-pressure clamp cell. A well-defined single excitation mode is observed. The softening of the excitation mode due to the applied pressure is clearly observed. From the analysis of the dispersion relations, it is found that an intradimer interaction decreases under hydrostatic pressure, while most interdimer interactions increase.}

\kword{KCuCl$_3$, spin gap, spin dimer, pressure, neutron inelastic scattering, magnetic excitations, dispersion relations, quantum phase transition, soft mode}

\begin{document}
\maketitle

For a system of isolated $S=1/2$ spin dimers with an antiferromagnetic (AF) intradimer exchange interaction $J > 0$, the triplet excitation is dispersionless and the excitation energy is given by ${\hbar}{\omega}(\mib q)=J$. When a three-dimensional interdimer exchange interaction $J'$ is switched on, excited triplets can hop to neighboring dimers due to the transverse component of the interdimer interaction $(J'/2)(S_i^+S_j^- + S_i^-S_j^+)$ so that the magnetic excitations become dispersive. With increasing $(J'/J)$, the dispersion range increases and the lowest excitation energy corresponding to the energy gap of the system decreases. At a quantum critical point $(J'/J)_{\rm c}$, the gap closes and the disordered ground state changes to the AF-ordered state \cite{Matsumoto1,Nohadani1}. The AF state stabilized for $(J'/J) > (J'/J)_{\rm c}$ is described by the coherent superposition of the singlet $|0,0\rangle$ and two triplet components $|1,{\pm}1\rangle$. Such a quantum phase transition (QPT) was first realized in TlCuCl$_3$ \cite{Goto1,Oosawa1,Rueegg1} by the application of hydrostatic pressure. The critical pressures determined by magnetization measurement and neutron scattering experiment are $P_{\rm c}=0.42$ kbar \cite{Goto1} and 1.07 kbar \cite{Rueegg1}, respectively. There is a discrepancy between the critical pressures obtained from macroscopic and microscopic measurements. 
 
\begin{figure}[htbp]
  \begin{center}
    \includegraphics[keepaspectratio=true,width=70mm]{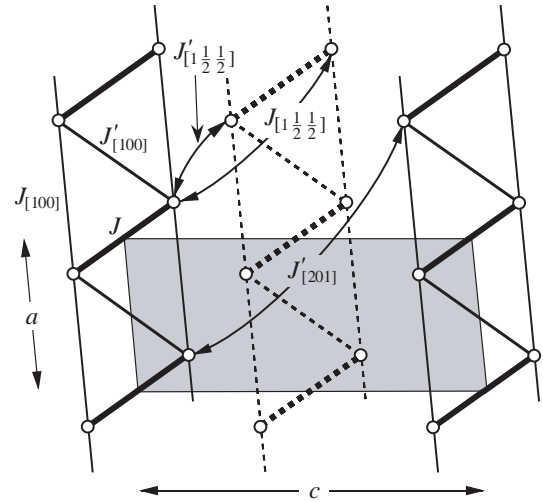}
  \end{center}
  \caption{Exchange interactions in the present systems. Double chains located at corners and the center of the chemical unit cell in the $bc$-plane are represented by solid and dashed lines, respectively. The shaded area shows the chemical unit cell in the $ac$-plane. $J$ denotes the intradimer interaction. $J_{[lmn]}$ and $J_{[lmn]}'$ denote interactions between dimers separated by the lattice vector $l{\mib a} + m{\mib b} + n {\mib c}$.}
  \label{fig:1}
\end{figure}

In this study, we investigate magnetic excitations in KCuCl$_3$ under hydrostatic pressure. The crystal structure (space group $P2_1/c$) of KCuCl$_3$ is the same as that of TlCuCl$_3$ \cite{Willett,Tanaka} and is composed of planar dimers of Cu$_2$Cl$_6$. The dimers are stacked to form infinite double chains parallel to the crystallographic $a$-axis. These double chains are located at the corners and center of the unit cell in the $bc$-plane. A strong AF exchange interaction with $J=4.34$ meV in the planar dimer of Cu$_2$Cl$_6$ dimerizes the spins of Cu$^{2+}$. As shown in Fig. 1, neighboring spin dimers couple along the $a$-axis and in the $(1, 0, {\bar 2})$ plane, in which the hole orbitals of Cu$^{2+}$ spread \cite{Kato1,Cavadini1,Mueller,Cavadini2,Kato2}. The lowest triplet excitation occurs at ${\mib Q}=(0, 0, 1)$ and its equivalent reciprocal points. The magnitude of the energy gap is ${\Delta}=2.67$ meV \cite{Kato1,Shiramura,Oosawa2}. Consequently, KCuCl$_3$ is magnetically characterized as an interacting spin dimer system.

KCuCl$_3$ undergoes a QPT under hydrostatic pressure, as observed in TlCuCl$_3$ \cite{Goto2}. The critical pressure obtained through magnetization measurement is $P_{\rm c}=8.2$ kbar \cite{Goto2}. An analysis of magnetic susceptibilities has revealed that the intradimer interaction $J$ decreases linearly under pressure, and that the sum of interdimer interactions increases with pressure. This causes the shrinkage of the gap and leads to the pressure-induced QPT. However, the pressure dependence of individual exchange interactions and the change of dispersion relations under pressure in KCuCl$_3$ have not yet been clarified, as well as in TlCuCl$_3$. To investigate these phenomena, we have performed neutron inelastic scattering experiments on KCuCl$_3$ under high hydrostatic pressure. In general, it is difficult to carry out neutron inelastic scattering under high pressure because of the small sample space and the attenuation of the neutron beam due to the thick pressure cell. In the present study, we used a newly designed cylindrical high-pressure clamp cell and obtained well-defined excitation spectra. In this paper, we report the results of our experiments.

\begin{figure}[htbp]
  \begin{center}
    \includegraphics[keepaspectratio=true,width=75mm]{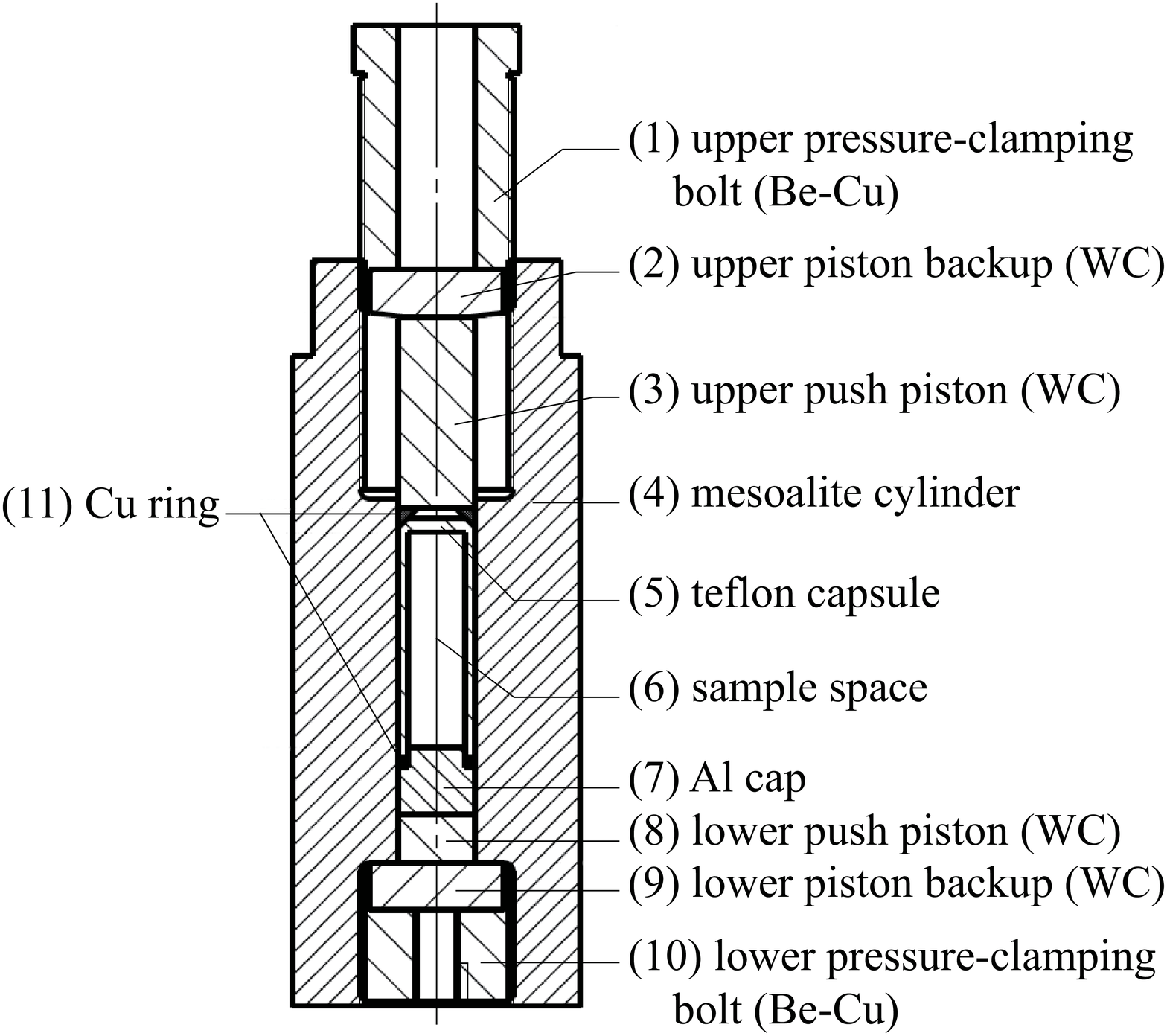}
  \end{center}
  \caption{Sectional view of the mesoalite hydrostatic high-pressure cell. The outer and inner diameters of the mesoalite cylinder are 29 mm and 10 mm, respectively.}
  \label{fig:2}
\end{figure}
Neutron inelastic scattering experiments were performed using the TAS-1 and LTAS spectrometers installed at JRR-3M in JAEA, Tokai. Figure 2 shows a sectional view of the cylindrical high-pressure clamp cell used for the present neutron scattering experiments. The cylinder with outer and inner diameters of 29 mm and 10 mm, respectively, is made of mesoalite (Al-Zn-Mg-Cu alloy), which is translucent for the neutron beam. The inner diameter of the sample capsule is 8.0 mm and the working height is 22 mm. This can accommodate a sample 7.8 mm in diameter and 16 mm in length. A KCuCl$_3$ single crystal prepared by the Bridgman method was shaped into a column 7.8 mm in diameter and 14 mm in length with the central axis parallel to the crystallographic $b$-axis. The KCuCl$_3$ sample was placed in a teflon capsule. A mixture of Fluorinert FC70 and FC77 was used as the pressure-transmitting medium. A hydrostatic pressure of $P=4.7$ kbar was applied at helium temperatures. The pressure was determined from the lattice constants of a NaCl crystal placed in the sample space. The sample was mounted in the cryostat with its $a^*$- and $c^*$-axes in the scattering plane. The constant-${\mib k}_f$ mode was adopted with a fixed final neutron energy $E_f$ of 14.7 meV and collimations were set as open$-80'-80'-80'$ on TAS-1. To investigate the low-energy excitation with a high energy resolution, inelastic scattering, in which the constant-${\mib k}_f$ mode was adopted with $E_f=3.5$ meV and collimations were set as open$-80'-80'-$open, was also carried out on LTAS. Sapphire and pyrolytic graphite filters were placed to suppress the background caused by high-energy neutrons and higher-order contaminations, respectively. The crystallographic parameters were determined as $a=3.918$ $\rm{\AA}$, $c=8.545$ $\rm{\AA}$ and $\beta=96.07^{\circ}$ at helium temperatures. The constant-${\mib Q}$ energy scan profiles were collected in the $a^*-c^*$ plane at $T=1.4$ K.

\begin{figure}[htbp]
  \begin{center}
    \includegraphics[keepaspectratio=true,width=75mm]{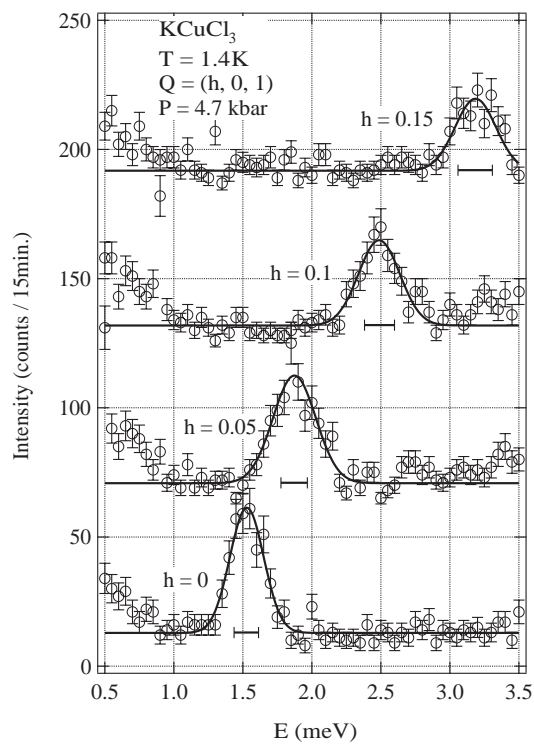}
  \end{center}
  \caption{Constant-${\mib Q}$ energy scan profiles for ${\mib Q}$ along $(h, 0, 1)$ with $0\leq h\leq 0.15$ in KCuCl$_3$ at $P=4.7$ kbar collected using the LTAS spectrometer. The solid lines are fits obtained using a Gaussian function. Horizontal bars denote calculated instrumental resolution widths.}
  \label{fig:3}
\end{figure}
Figure 3 shows profiles of constant-${\mib Q}$ energy scans for ${\mib Q}$ along $(h, 0, 1)$ collected using the LTAS spectrometer. A well-defined single excitation mode can be observed in almost all scans. On the basis of previous studies \cite{Kato1,Cavadini1,Cavadini2,Kato2}, the observed excitations can be identified as one-triplet excitations. No splitting of the excitation spectrum due to anisotropy energy was observed. The scan profiles were fitted with a single Gaussian function to evaluate the excitation energy, as shown by the solid lines in Fig. 3. 
The horizontal bars in Fig. 3 denote the calculated resolution widths. Almost all peaks have widths close to the resolution limit. 
The dispersion relations $\omega({\mib Q})$ in KCuCl$_3$ obtained at $P=4.7$ kbar for $\mib Q$ along $(h, 0, 1)$, $(0, 0, 2h+1)$, $(h, 0, 2h+1)$ and $(h, 0, 2h-1)$ are summarized in Fig. 4. The dashed lines in Fig. 4 denote the fits of the dispersion curves at ambient pressure \cite{Kato1,Cavadini1,Mueller,Cavadini2,Kato2} obtained using eq. (1) with the exchange parameters listed in Table I. Softening of the triplet mode is clearly observed at all $\mib Q$ values. The excitation gap for $P=4.7$ kbar corresponding to the lowest excitation energy was evaluated as ${\Delta}=1.52$ meV. Figure 5 summarizes the excitation gaps obtained by the present study and previous magnetization \cite{Goto2,Shiramura,Oosawa2} measurements at various pressures. The present result of ${\Delta}=1.52$ meV for $P=4.7$ kbar is in accordance with others.   

\begin{figure}[htbp]
  \begin{center}
    \includegraphics[keepaspectratio=true,width=80mm]{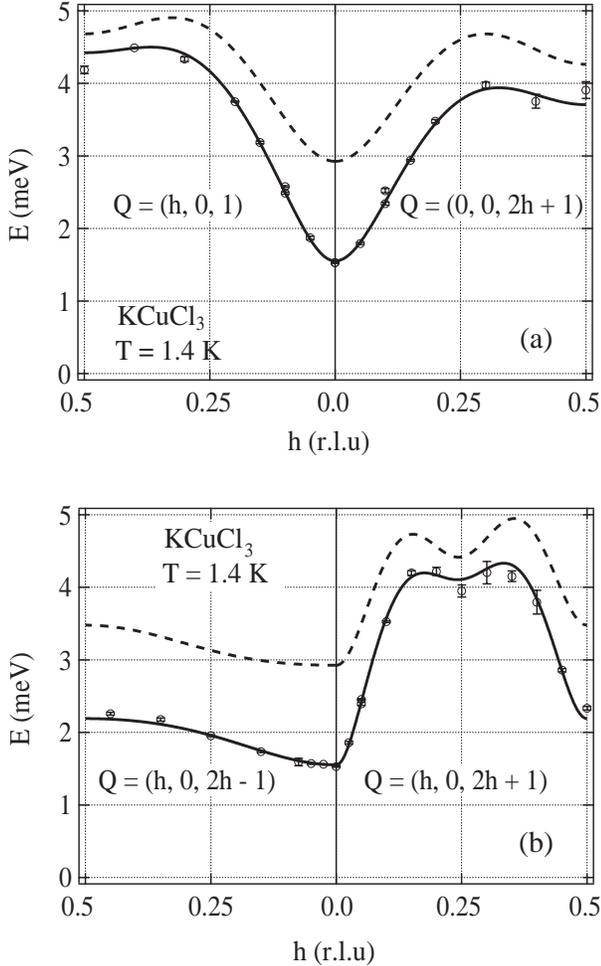}
  \end{center}
  \caption{Dispersion relations $\omega(\mib Q)$ in KCuCl$_3$ measured at $P=4.7$ kbar for $\mib Q$ along (a) $(h, 0, 1)$ and $(0, 0, 2h+1)$, and (b) $(h, 0, 2h+1)$ and $(h, 0, 2h-1)$. Solid and dashed lines are the dispersion curves at $P=4.7$ kbar and ambient pressure, respectively, calculated using eq. (1) and the exchange parameters listed in Table I.}
  \label{fig:4}
\end{figure}

As shown in Fig. 4, the dispersion relations at ambient pressure and $P=4.7$ kbar are generally similar to each other. However, the average of the dispersion decreases with increasing pressure, which indicates a decrease in the intradimer interaction $J$ with increasing pressure. On the other hand, the bandwidth of the dispersion for $P=4.7$ kbar is approximately 1.5 times larger than that for ambient pressure, which indicates an increase in the interdimer interactions with pressure. These pressure dependences of the intradimer and interdimer interactions are consistent with the results obtained by the magnetization measurements \cite{Goto2}. The pressure dependence of the dispersion relations in KCuCl$_3$ differs from that in the spin-Peierls system CuGeO$_3$, in which the gap increases with pressure, while the dispersion range decreases \cite{Nishi}.
\begin{table}[tb]
\caption{Intradimer and effective interdimer exchange interactions in KCuCl$_3$ in meV units. Exchange parameters for ambient pressure were taken from refs. \citen{Cavadini1,Mueller,Cavadini2,Kato2}.
\label{table1}}
\begin{center}
\begin{tabular}{ccccc} \hline
$P$ [kbar] & $J$ & $J^{\rm eff}_{[1\,0\,0]}$ & $J^{\rm eff}_{\left[1\,\frac{1}{2}\,\frac{1}{2}\right]}$ & $J^{\rm eff}_{[2\,0\,1]}$ \\ \hline
ambient & $4.34$ & $-0.21$ & $0.28$ & $-0.45$ \\ \hline
$4.7$ & $3.65$ & $-0.28$ & $0.45$ & $-0.42$ \\ \hline
\end{tabular}
\end{center}
\end{table} 
\begin{figure}[htbp]
  \begin{center}
    \includegraphics[keepaspectratio=true,width=80mm]{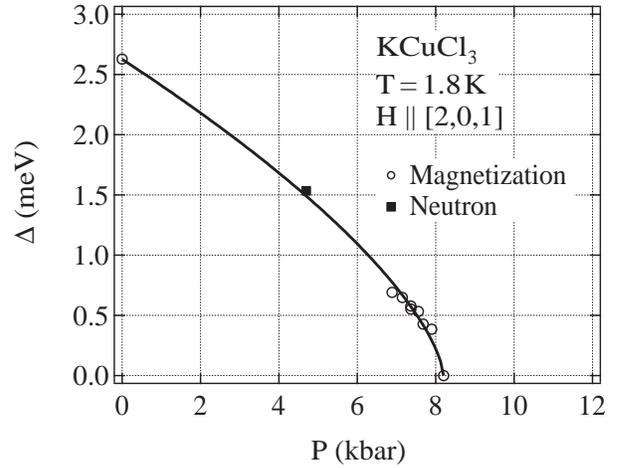}
  \end{center}
  \caption{Pressure dependence of the excitation gap $\Delta$ in KCuCl$_3$. The solid line serves as a visual guide.}
  \label{fig:5}
\end{figure}

Next, we analyze the dispersion relations using the effective dimer interactions for the propagation of triplet excitations \cite{Cavadini1,Mueller,Cavadini2,Kato2,Cavadini3,Oosawa3}. Since there are two different dimers per chemical unit cell, the triplet excitations have two branches $\omega_{\pm}({\mib Q})$. Within the framework of the random phase approximation, the dispersion relations in the present system are expressed as\, \cite{Cavadini1,Mueller,Cavadini2,Kato2,Cavadini3,Oosawa3}
\begin{equation}
\label{RPA}
{\hbar}\omega_{\pm}({\mib Q})=\sqrt{J^2+ 2JJ^{\rm eff}_{\pm}({\mib Q})}\,,
\end{equation}
where
\begin{eqnarray}
\label{omega}
J^{\rm eff}_{\pm}({\mib Q}) = \left[J_{[100]}^{\rm eff}{\cos}(2{\pi}h) + J_{[201]}^{\rm eff} {\cos}\{ 2{\pi}(2h+l)\}\right] \nonumber \\ 
{\pm}\, 2J_{\left[1\frac{1}{2}\frac{1}{2}\right]}^{\rm eff} {\cos}\{{\pi}(2h+l)\}{\cos}({\pi}k)\,.\hspace{1.6cm} 
\end{eqnarray}
In the above equations, $J$ denotes the main intradimer exchange interaction and $J_{[lmn]}^{\rm eff}$ denotes the effective interaction between dimers separated by a lattice vector $l{\mib a} + m{\mib b} + n{\mib c}$ and is expressed as
\begin{equation}
J^{\rm eff}_{[lmn]} = \frac{1}{2}\left(MJ_{[lmn]} - M'J'_{[lmn]}\right),
\end{equation}
where $J_{[lmn]}$ and $J_{[lmn]}'$ are shown in Fig. 1. The factors $M$ and $M'$ are the numbers of identical exchange paths for the same lattice vector between dimers. For $J_{[100]}^{\rm eff}$, $M=2$ and $M'=1$, and for the others, $M=M'=1$. The effective dimer interaction $J_{[lmn]}^{\rm eff}$ corresponds to the hopping amplitude of the triplet excitation between dimers. Under the present experimental condition, i.e., $\mib Q$ in the $a^*-c^*$ plane, the $\omega_{-}(\mib Q)$ branch vanishes and only the $\omega_{+}(\mib Q)$ branch can be observed. The solid lines in Fig. 4 denote the calculated results obtained with the exchange parameters shown in Table I. The experimental dispersion curves can be reproduced well by the fitting. We also fitted the dispersion curves, including $J_{[200]}^{\rm eff}$ and $J_{[0\frac{1}{2}\frac{1}{2}]}^{\rm eff}$, and found that their contribution is negligible.

The intradimer interaction $J$ decreases significantly with pressure. From previous magnetization measurements \cite{Goto2}, it was found that $J$ exhibits linear dependence on pressure $P$ as $J(P)=J(0)\left(1-P/27.0\right)$ in units of kbar. Using this relation, we obtain $J=3.58$ meV at $P=4.7$ kbar. This $J$ is in agreement with the present result of $J=3.65$ meV. It is considered that the bond angle of the intradimer exchange path $\rm{Cu^{2+}-Cl^{-}-Cu^{2+}}$ approaches 90$^{\circ}$ with increasing hydrostatic pressure, so that the ferromagnetic contribution to the intradimer interaction increases and the total AF intradimer interaction decreases. 

As shown in Table I, the most dominant interdimer interaction $J'_{[201]}=-(1/2)J^{\rm eff}_{[201]}$ acting between the next-nearest dimers in the $(1, 0, {\bar 2})$ plane scarcely varies with pressure. On the other hand, the effective interdimer interactions $J_{[100]}^{\rm eff}$ and $J_{[1\frac{1}{2}\frac{1}{2}]}^{\rm eff}$ acting between the nearest dimers along the $a$-direction and in the $(1, 0, {\bar 2})$ plane exhibit a significant increase with pressure. The enhancement in the effective interdimer interactions $J_{[100]}^{\rm eff}$ and $J_{[1\frac{1}{2}\frac{1}{2}]}^{\rm eff}$ is attributed to the contraction of the interdimer distance. M\"{u}ller and Mikeska \cite{Mueller} have evaluated the individual interdimer interactions at ambient pressure in KCuCl$_3$ by applying a cluster series expansion to the analysis of the dispersion relations. According to their results, $J'_{[100]}$, $J_{[1\frac{1}{2}\frac{1}{2}]}$ and $J'_{[201]}$ are dominant in the interdimer interactions shown in Fig. 1. Therefore, we infer that the increase in $J'_{[100]}$ and $J_{[1\frac{1}{2}\frac{1}{2}]}$ contributes to the increase in $J_{[100]}^{\rm eff}$ and $J_{[1\frac{1}{2}\frac{1}{2}]}^{\rm eff}$ with pressure.

Matsumoto {\it et al.} \cite{Matsumoto1} theoretically investigated the evolution of magnetic excitations in the present system across the critical pressure $P_{\rm c}$. They predicted that after the softening of the triply degenerated triplet mode at $P_{\rm c}$, the excitations are reorganized into two gapless phase modes and one gapped amplitude mode. The amplitude mode, which has been disregarded in conventional antiferromagnets, corresponds to the oscillation of the length of ordered moments. The excitation energy of the amplitude mode increases rapidly with increasing pressure, while its intensity decreases rapidly so that it vanishes at high pressures. The observation of the novel amplitude mode is of great interest. We are now improving the high-pressure cell for neutron inelastic scattering to reach a hydrostatic pressure of 10 kbar. 

In conclusion, we have presented the results of neutron inelastic scattering experiments on the gapped spin system KCuCl$_3$ under a hydrostatic pressure of 4.7 kbar, which was applied using a newly designed cylindrical high-pressure clamp cell made of mesoalite. A well-defined single excitation mode was observed. The dispersion relations along four different directions in the $a^*-c^*$ plane were obtained as shown in Fig. 4. The excitation energy decreases at all ${\mib Q}$ values under hydrostatic pressure. Using the random phase approximation, we have analyzed the dispersion relations and obtained the intradimer and effective interdimer interactions. It was found that the intradimer interaction decreases under hydrostatic pressure and the effective interactions between the nearest dimers along the $a$-direction and in the $(1, 0, {\bar 2})$ plane increase. These pressure dependences of the exchange interactions give rise to the observed pressure-induced QPT from a gapped ground state to an AF state in KCuCl$_3$.

The authors thank M. Matsuda for useful discussions. This work was supported by a Grant-in-Aid for Scientific Research from the Japan Society for the Promotion of Science and a 21st Century COE Program at Tokyo Tech ``Nanometer-Scale Quantum Physics'' from the Ministry of Education, Culture, Sports, Science and Technology of Japan. A. O. was supported by the Saneyoshi Scholarship Foundation and the Kurata Memorial Hitachi Science and Technology Foundation.

\end{document}